\documentclass[runningheads]{llncs}
\usepackage[utf8]{inputenc} 
\usepackage{graphicx}
\usepackage[caption=false]{subfig}
\usepackage{url}
\usepackage{booktabs}       
\usepackage{todonotes}
\newcommand{\system}{\textsc{Tedric}}
\newcommand{\systemExplanation}{Talk Exercise Designer using Random Inspirational Combinations}

\begin{document}
\title{Automatically Generating Engaging Presentation Slide Decks}
\author{Thomas Winters\inst{1}\orcidID{0000-0001-7494-2453} \\ \and
Kory W. Mathewson\inst{2}\orcidID{0000-0002-5688-6221}}
\institute{KU Leuven, Leuven, Belgium, \\
\email{thomas.winters@cs.kuleuven.be} \and
University of Alberta, Alberta, Canada \\\
\email{korymath@gmail.com}}

\maketitle 
\begin{abstract}
    Talented public speakers have thousands of hours of practice.
    One means of improving public speaking skills is practice through improvisation, e.g. presenting an improvised presentation using an unseen slide deck.
    We present \system, a novel system capable of generating coherent slide decks based on a single topic suggestion.
    It combines semantic word webs with text and image data sources to create an engaging slide deck with an overarching theme.
    We found that audience members perceived the quality of improvised presentations using these generated slide decks to be on par with presentations using human created slide decks for the \textit{Improvised TED Talk} performance format.
    \system\ is thus a valuable new creative tool for improvisers to perform with, and for anyone looking to improve their presentation skills.
  
\end{abstract}
\section{Introduction}

Public speaking is difficult due to many psychosocial factors:
40\% of surveyed adults feel anxious about speaking in front of an audience \cite{wilbur1981,doi:10.1080/08824096.2012.667772}.
Practice and experience can make significant improvements to speech anxiety levels \cite{lee2014study}.
However, the overpracticing of a talk can make the delivery stiff and rehearsed.
Improvisational theater has been shown to be effective in developing skills for complex social interactions \cite{watson2011perspective,king1956comparison}.
Giving improvised presentations is thus a useful exercise for enhancing presentation skills and overcoming public speaking anxiety.
The exercise itself is already performed by improvisational theater and comedy groups all around the world in several variants, such as in the form of an \textit{``Improvised TED Talk''} \cite{mathewson-ted-talks}.
This work presents the novel idea of automatically generating presentations suitable for such an improvisational speaking practice exercise, as well as for improvisational comedy performance.

One of the difficulties in the improvised presentation format is that the slide deck is created prior to performance, while in improvisational comedy, the audience provides suggestions to shape the show.
The premade slide deck
will thus not align with the suggestion, but instead just has \textit{``curated random''} slides \cite{mathewson-ted-talks}.
Another difficulty is learning how to build constructive slide decks for another performer to present.
Creating such a slide deck requires time and attention, which can be costly and does not scale well when doing a large quantity of performances.
To solve these problems, we present \system\ (\textit{\systemExplanation}), a co-creative generative system that is capable of creating engaging slide decks for improvised presentations.
It is designed to fulfill a specific task of an improviser for a particular format, being the design of an improvised TED-talk slide deck \cite{mathewson-ted-talks}.
In the improvised presentation format, one performer usually designs a slide deck for the other performer beforehand.
Previous systems in this domain have experimented with fulfilling the presenter role, e.g. by writing text scripts, creating comedic performers, and all-round improvisational comedy agents \cite{samim-ted-rnn,knight-standup-robot,mathewson2017improvised}.
\system\ on the other hand focuses on the task of the other performer, who creates the slide deck for the speaker.
The system also differs from previous work in slide deck generation \cite{ppt-gen-academic,ppt-gen-review,Sravanthi2009SlidesGenAG} in that rather than summarizing text into slides, or converting outlines to well-designed slides, it actively creates its own presentation story and chooses the content itself.
This system is thus a good example of co-creation between humans and machines, as both have responsibility in the creative process of performing an improvised presentation.
Thanks to the speed advantage \system\ has over its human counterpart, it can generate a slide deck based on a suggestion from a live audience on the spot.
Another advantage \system\ has, is that it follows certain good design guidelines, something which individuals new to this format have struggled to learn quickly \cite{mathewson-ted-talks}.
  
\section{Slide deck generator model}
The \system\ system is composed of several linked generators, specified in a presentation schema.
In this paper, we focus on the \textit{Improvised TED talk} presentation schema, which  is meant to give comedic suggestions to spark the presenter's creativity during the presentation.
However, it is easy to create similar schemas for more serious presentations (e.g. an improvised \textit{Pecha Kucha} \cite{PechaKucha}) by taking the subset of slide generators that do not use comedic content sources.

\begin{figure}
  \centering
    \includegraphics[width=\textwidth]{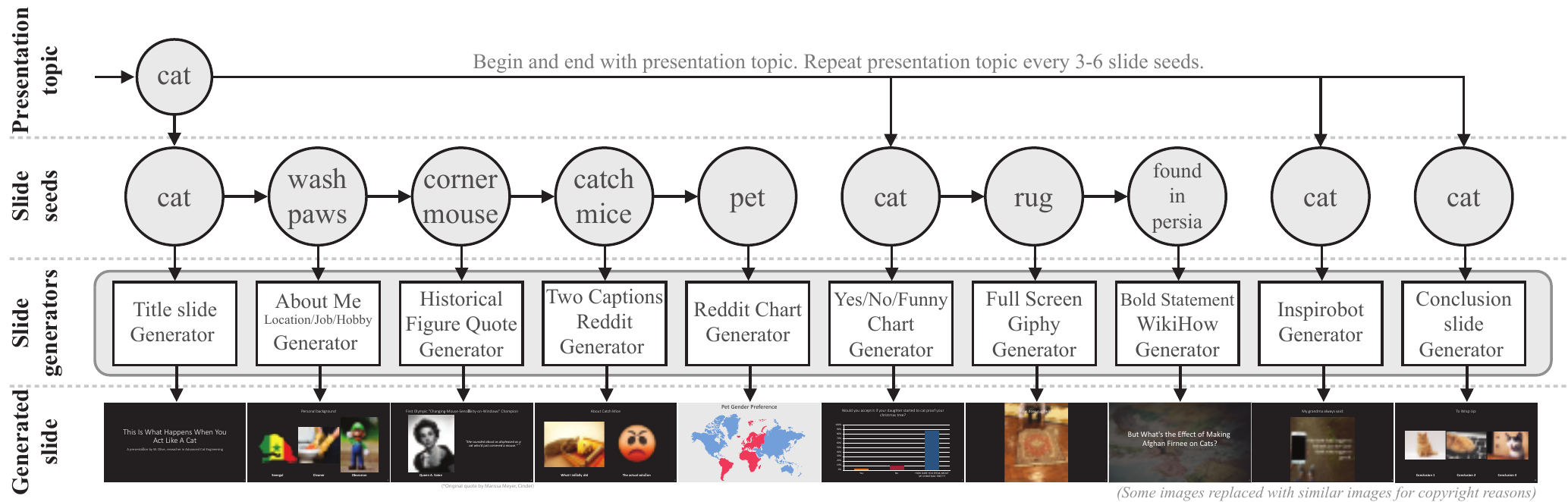}
  \caption{\system\ system pipeline}
  \label{fig:pipeline}
\end{figure}

\subsection{Schemas}
The templates and schemas technique is a commonly used approach for generating content using certain constraints \cite{JAPE,STANDUP-Construction,venour:1999,winters2018jokegeneration}.
In this project, the presentation schema specifies a slide seed generator, a set of slide generators, slide generator weight functions and frequency limitations on slide generator types.
The slide generators themselves also use a different kind of schema to generate their content.
Such a slide generator schema specifies a set of content sources to fill placeholders in a specific PowerPoint slide template, resulting in a coherent slide.

\subsection{Slide Seed Generator}
A seed generator produces a list of slide seeds based on the overarching main presentation topic, provided by the audience.
The default seed generator achieves this by randomly walking over links in ConceptNet, a semantic network of related words \cite{conceptnet}.
For example, it could find that \textit{cat} is related to \textit{rug} because cats can be found laying on rugs (Figure \ref{fig:pipeline}).
If no useful link is found for a certain slide, it backtracks and looks at the slide seed before the previous slide seed.
The default seed generator constrains the first and last slide to be about the given presentation topic, as well as making this main topic return every three to six slides.
It thus ensures that the given main topic is clearly present in the presentation, while still allowing some deviations to related topics.

\subsection{Slide Generator}
\label{sec:slide-generator}

Slide generators use a given seed to generate a logical, themed slide for the presentation slide deck.
During slide deck generation, the presentation schema iteratively picks one of its slide generators using their weight functions.
These weight functions calculate the weight of a slide generator based on the current position of the slide to be generated and on how often similar slide generators can occur in one presentation.
For example, the weight of the \textit{``Title Slide Generator''} peaks for the first slide, and has a weight of zero everywhere else.
Similarly, the several \textit{``About Me"} and \textit{``History''} slide generators prefer being in one of the first slides after the title slide, and thus have higher weights when choosing a slide generator for these positions.
Each slide generator also has several associated tags.
These tags are used to limit the number of certain slide types occurring in the same slide deck.
For example, at most one anecdote slide is allowed, and no more than 20\% of the total number of slides can be about a quote.
For every slide, the presentation schema thus first creates a list of slide generators that can still be used given the slide generators used for other slides.
It then calculates the weight for all these slide generators by giving the slide number and the total number of slides as arguments to their weight functions.
It then picks the slide generator using roulette wheel selection on the weights of these slide generators.

After picking a slide generator, this generator then creates a slide based on a given seed, using its slide generation schema.
To achieve this, it passes the seed to all content sources specified in this template.
For example, using the seed \textit{``rug''}, the \textit{``Full Screen Giphy Generator''} passes this seed to Giphy to find a related gif, and to a title content source to generate a slide title using this seed word (Figure \ref{fig:pipeline}).

The system currently contains 26 slide generators.
They can be grouped in several different main categories based on the type of slide they produce, which are listed below.

\subsubsection{Title Slide.}

The title slide generator creates a slide with a generated title about the main topic, based on a large set of templates.
The seed is often transformed to a related action by using WikiHow (see Section \ref{sec:wikihow-explanation}).
The slide also contains a subtitle with a generated presenter name (if none is provided) and a generated scientific sounding field (e.g. \textit{Applied Cat Physics}) (Figure \ref{fig:title-slide}).

\subsubsection{About Me Slide.}

There are several different slide generators introducing the speaker.
They all generate texts and images visualizing the accompanied text, and can be about a location, a country, a job, a book and/or a hobby related to the presenter (Figure \ref{fig:about-me-slide}).

\subsubsection{History Slide.}

The system contains several generators that make up history related to the given seed.
The \textit{``Historical Figure Quote''} picks an old picture (from Reddit), generates a name (using a context-free grammar), invents an achievement (using a WikiHow action and templates) and finds a related quote (using GoodReads) to create a fake historical figure related to the seed (Figure \ref{fig:historic-figure-slide}).
There are also other slide generators using vintage pictures with captions relating to time.
These history slides are used for adding depth and narrative to the generated presentation topic.

\subsubsection{Full Screen Image Slide.}

Breaking up the slideshow with a full screen image is a great way for providing freedom as well as for grabbing attention.
There are several variants of large image slide generators employed in \system, such as using Google Images to create a descriptive slide, where the speaker has a lot of freedom and breathing space to talk.
Another variant uses a large animated gif and a title pointing the attention to the gif.

\subsubsection{Statement Slide.}

Since the slide decks are meant for \textit{Improvised TED talks}, we want them to feel as if they provide the audience with a take away message.
To achieve this, we created several templates for bold statements, call to actions and anecdote prompts (Figure \ref{fig:anecdote-slide}).
The system also uses several online sources (such as InspiroBot and GoodReads) to provide inspirational quotes.

\subsubsection{Multiple Captioned Images Slide.}
Some slide generators use multiple images on one slide, with randomly picked captions that belong together underneath (e.g. \textit{``Expectation''} \& \textit{``Reality''}).
We found that these slides work best if the last captioned image uses content source providing odd images, as they help the presenter with creating a strong punchline if done correctly.

\subsubsection{Chart Slide.}
\begin{figure}
\subfloat[Title slide about \textit{``phone''}\label{fig:title-slide}]{%
      \includegraphics[width=.45\linewidth]{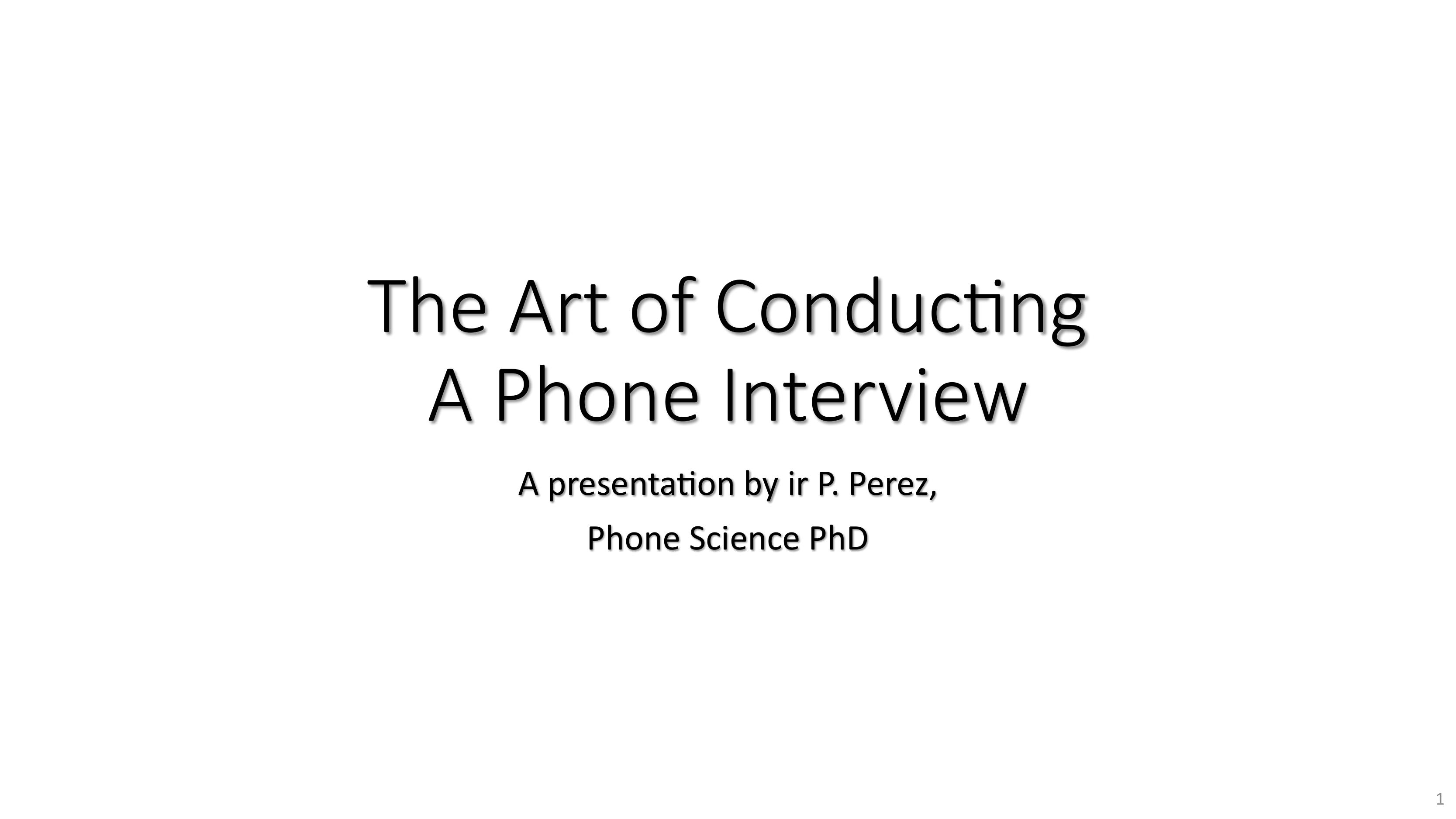}
     }
     \hfill
\subfloat[About Me (Location-Job-Hobby) slide\label{fig:about-me-slide}]{%
  \includegraphics[width=.45\linewidth]{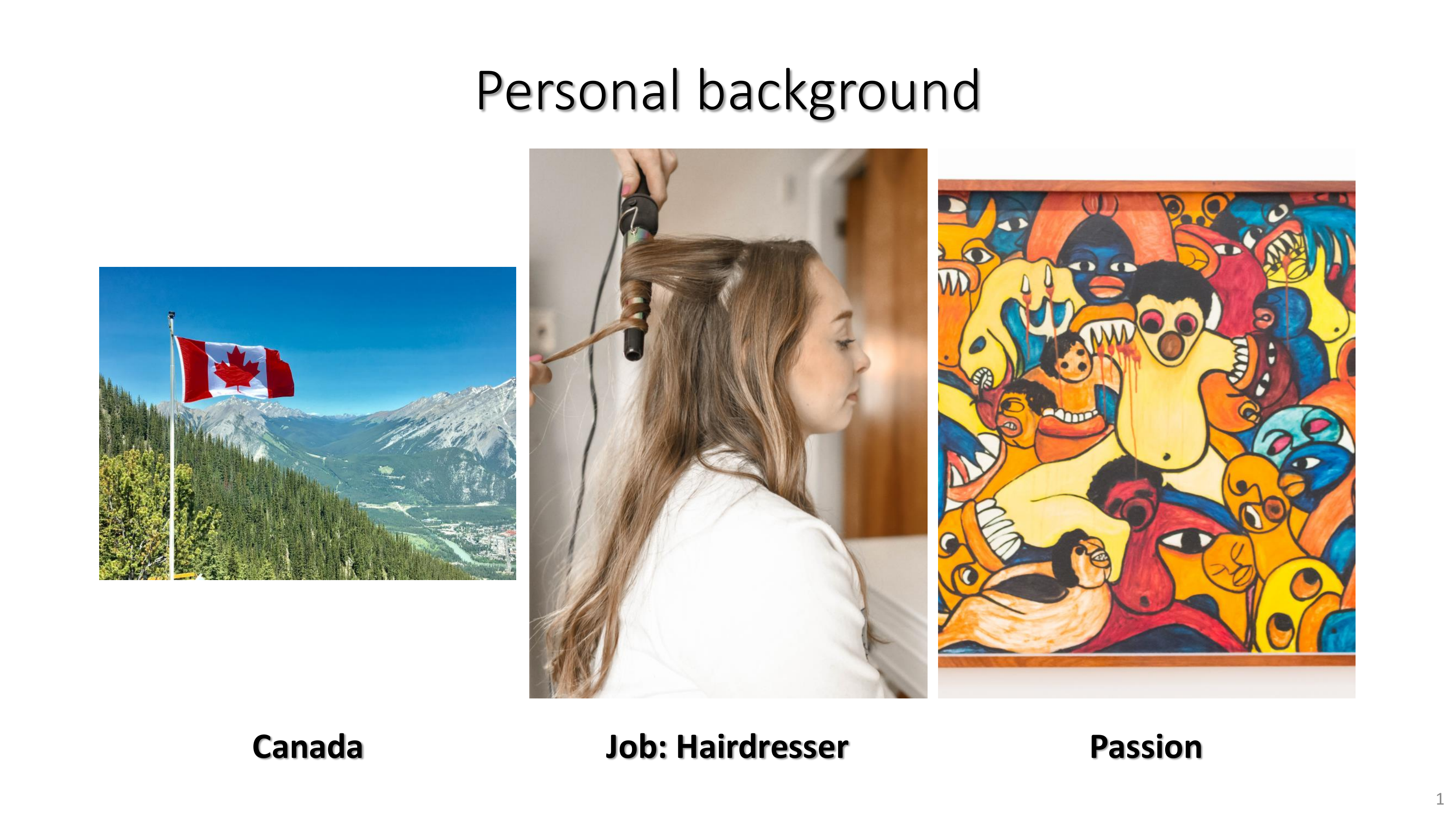}
     }

\subfloat[Historical Figure slide about \textit{``car''}\label{fig:historic-figure-slide}]{%
  \includegraphics[width=.45\linewidth]{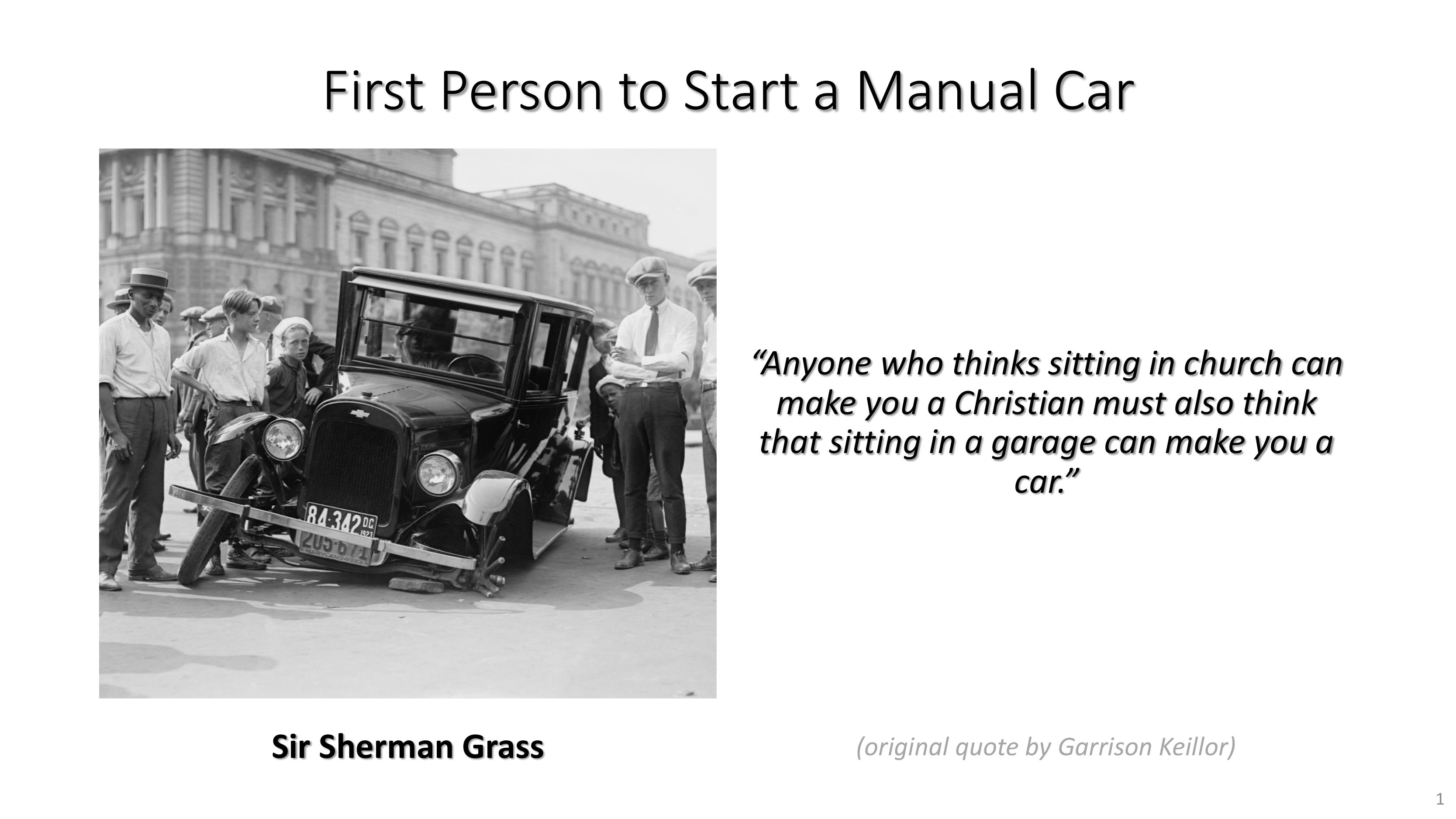}
     }
     \hfill
\subfloat[Location Pie Chart slide about \textit{``room''}\label{fig:pie-slide}]{%
  \includegraphics[width=.45\linewidth]{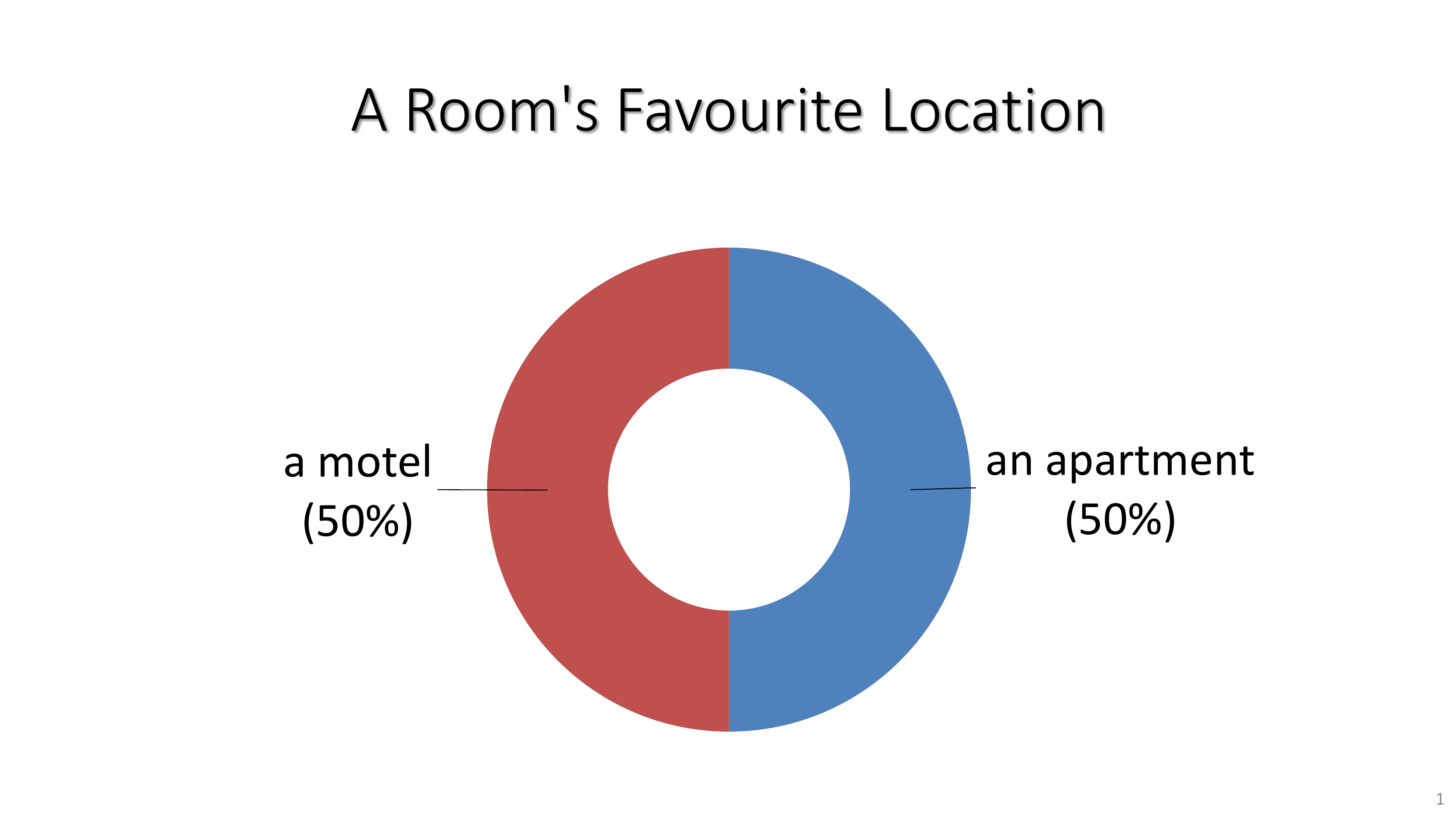}
     }

\subfloat[Anecdote slide about \textit{``fish''}\label{fig:anecdote-slide}]{%
  \includegraphics[width=.45\linewidth]{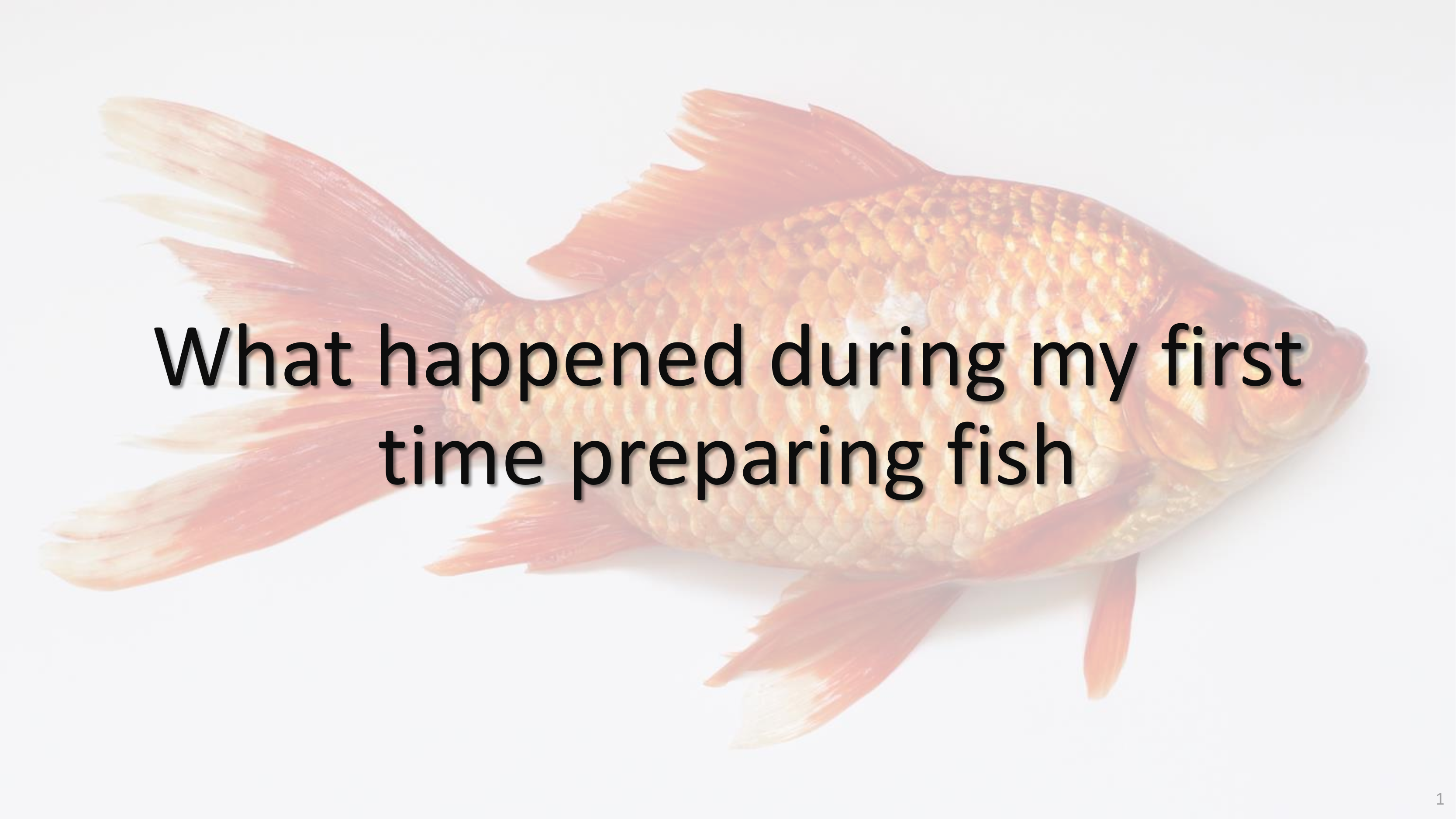}
     }
     \hfill
\subfloat[Yes/No/Funny Chart slide about \textit{``cat''}\label{fig:histogram-slide}]{%
  \includegraphics[width=.45\linewidth]{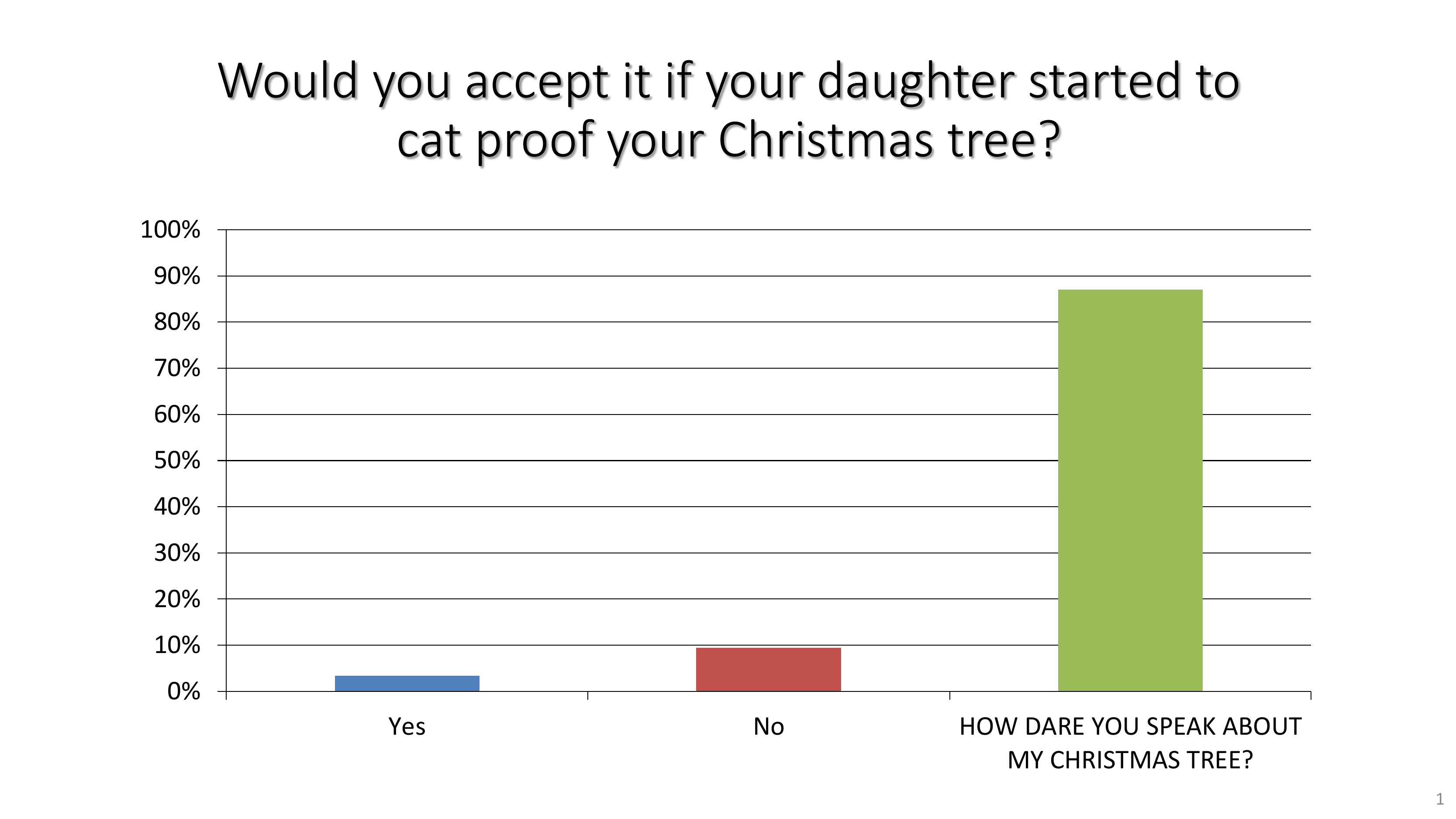}
     }

\caption{Possible outcome examples of slide generators}
\label{fig:slide-generators}
\end{figure}

We implemented several ways of creating charts using generated data, as these usually give the presentation the perception of having more scientific foundation.
One chart generator uses histograms and pie charts about yes/no questions.
It generates a question about the slide seed using templates, and a funny third answer based on a large collection of textual templates (Figure \ref{fig:histogram-slide}).
The data generator prefers generating high values for the funny answer.
Similarly, ConceptNet provides a good source of related locations that can be used in a pie chart or histogram (Figure \ref{fig:pie-slide}).
Another type creates scatter plots by adding noise to basic mathematical functions (e.g. $y=x^2$ and $y=log(x)$).
The labels of the axis are then found by using links in ConceptNet.
A third type uses subreddits about data visualization for finding interesting related charts.

\subsubsection{Conclusion Slide.}
Conclusion slides use a similar structure to the multiple captioned image slides.
These slides usually start with a descriptive image about the main topic and have an odd image at the end, in order to finish the show with a punch.

\subsection{Content Sources}
\label{sec:content-sources}
Content sources are used in slide generators to provide them with either texts or images.
The content sources produce this content based either on a suggestion or at random.
Slide generators pass the seeds they receive from the presentation schema down to all of their content sources.
Each content source then decides whether or not to use this seed.
For example, if the content source uses a search engine, it might fall back to picking content at random in case the search engine does not return any results.
Some content sources (e.g. a context-free grammar for generating full names) might not support using the seed at all, and thus always return completely random content.
Ignoring the seed and picking content at random every once in a while can have a positive effect as it might provide more variety in a presentation.
This is especially true when the slide seed generator is not able to find a lot of meaningful related words, e.g. for an uncommon, niche word.

The system uses two types of content sources: image sources and text sources.
These use search systems, context-free grammars, templates and combinations of the three.
One advantage search systems have over pure context-free grammars and templates is that they are usually larger, contain more diverse content and are frequently updated.
This gives rise to more surprises to both the creators as well as the users of the system, as new entries are usually constantly being added to these sources, generating different content every time it is used.
Since such sources are harder to predict, it is important to identify the style and flavour of the elements in the search system or corpus in order to successfully use it in a slide generator schema.
For example, stock photo websites usually give more neutral images, whereas certain subreddits might give funny or cute images.
Sources of similar flavour can then be combined into composite content sources, to increase the variety a slide generator can produce.
For example, the Google Images data source can be combined with the stock photo data source, as both provide neutral images.
The slide generators then use these content sources in places where they need data in such a specific style and flavour.

\subsubsection{Image Sources}
The system uses a large variation of image sources.
In order to find descriptive pictures, it uses Google Images and stock photo websites.
The system also uses images from Inspirobot\footnote{\url{http://inspirobot.me/}}, a bot that generates nonsensical inspirational quotes, and frames them using a thought-provoking background.
We also built our own data generation system for the analytical slide generators.
This data source is used to insert random data into charts using labels created by text sources.
Reddit\footnote{\url{https://reddit.com}} is also a great source of images, as a lot of its subreddits usually contain images in a particular style or flavour.
Images from Reddit are usually intrinsically interesting, as users tend to have good reasons for thinking the picture is worth posting on Reddit.
The number of upvotes for a Reddit post generally gives a good indication for the quality of the image as well.
We specifically use subreddits about odd, punchline images (e.g. \textit{r/hmmm}, \textit{r/wtfstockphotos}, \textit{r/eyebleach}), gifs (e.g. \textit{r/gifs}, \textit{r/nonononoYES}), historic and vintage pictures (e.g. \textit{r/OldSchoolCool}, \textit{r/TheWayWeWere}, \textit{r/ColorizedHistory}), charts and other data visualisations (e.g. \textit{r/dataisbeautiful}, \textit{r/funnycharts}), book covers and outside pictures.
The Reddit gif source is combined with the Giphy\footnote{\url{https://giphy.com/}} source to create a composite gif data source.

\subsubsection{Text Sources}
We employed several types of text sources in this project.
The first type utilizes search engines to find texts related to a slide seed.
As such, the system uses GoodReads\footnote{\url{https://www.goodreads.com/}} as a source for related quotes, to be used on slides about historic people or for large statements.
\label{sec:wikihow-explanation}
Another such search engine is WikiHow\footnote{\url{https://wikihow.com}}, which we use to find human actions related to a certain seed, or related to other actions.
This is achieved by searching on WikiHow for the given input and looking at the titles of the found articles.
Considering WikiHow article titles follow structures such as \textit{``How to ...''} or \textit{``5 ways to ...''}, it is trivial to extract actions related to the input from these.
For example, searching for \textit{``cat''} on WikiHow finds articles such as \textit{``How to Pet a Cat''}, which will lead to the action \textit{``to pet a cat''}, which can then for example be inserted into a template (e.g. ``\texttt{Why We All Need to \{seed.wikihow\_action.title\}}'') to form the talk title \textit{``Why We All Need to Pet a Cat''}.

For creating controllable text content sources with high variation, we built our own text generation language on top of Tracery, a text generation language using context-free grammars \cite{tracery}.
The extensions allows us to specify external variables with custom functions in the template texts.
These custom functions can be used on the externally specified variables, and thus transform them into right forms, or chain other functions.
They can also be used as a predicate (e.g. checking if the variable is a noun).
The transformation functions are used to change the form of the variable (e.g. verb conjugation, pronouns or string casing) or finding related texts (e.g. related actions using WikiHow or related concepts using ConceptNet \cite{conceptnet}).
This extension is implemented by first making Tracery generate text which might have external variables and functions specified between curly brackets.
After this generation, the system checks if there are missing external variables, failed predicate checks on these variables and if they are transformable using the specified functions.
If all conditions are satisfied, the variable is filled in, transformed and inserted in the textual template.
Otherwise, the generator keeps trying until all conditions are  met.
The extended generative grammar language allows us to specify a large number of generators, such as generators for historic people names, for bold and inspirational statements, talk (sub)titles, related sciences, slide title variations and punchline chart question answers.

\subsubsection{Tupled Sources}

For specifying certain slide generators, we required tupled data sources, i.e. data sources that inspire each other, or that are generated at the same time.
For example, when generating a job, we want to generate an image based on the generated job function.
When generating slides with two captions, these captions should sometimes also relate to each other (e.g. \textit{``good''} \& \textit{``bad''}, \textit{``what I initially did''} \& \textit{``actual solution''}).
For this, the system thus also allows the generation of content using tupled data source.

\subsection{Parallelisation}

Since the slide decks are meant to be generated in real-time in front of an audience, we need the generator to finish within a reasonable time frame.
To achieve this, the generator has the option of generating all slides at once in parallel.
This decreases the influence of slow (e.g. online, rate-limited) data sources.
However, naively generating all slides in parallel causes issues with the validity of the slide deck, as the generator has several constraints that must hold between generated slides, such as limited duplicate images and restrictions on maximum tag occurrences of the used slide generators.
To solve this, \system\ generates the slide seeds and selects some slide generators. 
It then fires all the slide generators at once without giving them any knowledge about other slides, since there are no other slides yet.
After all initial slides are generated, \system\ sequentially checks every slide, increasing its saved knowledge about the slides after viewing every next slide.
If a slide is breaking any of the constraints, its slide number is stored for the next generation round.
After viewing all slides, a new slide generation round is fired, but this time with knowledge from all valid, generated slides.
This repeats itself until no new slide breaks any constraints.
At the end of this process, the generated slides respect the constraints specified by the presentation schema similar to slides generated in a serial way.

\subsection{Code}

The code, documentation and link to an online demo of \system\ is available on
\url{https://korymath.github.io/talk-generator/}.

\section{Evaluation}

We evaluated the system in several different ways.
First, we made volunteers present improvised presentations and formally give feedback to all presentations.
This feedback is then analysed to compare the quality of human created slide decks versus generated slide decks.
Second, we evaluated the time performance of the full slide generation process in a context similar to the expected use case context, to evaluate if it completes within a reasonable time frame.
Third, we informally evaluated the system by training some performers using generated slide decks, and making them perform the format on stage using slide decks generated during the performance.

\subsection{Quantitative results}
To evaluate the quality of our system, we compared the perceived quality of generated slide decks with slide decks created by humans by performing them in the improvised presentation format with eight volunteers.
In total, six talks were given.
We generated three slide decks on the spot based on the audience suggestion for the presentation topic, and used three slide decks created by three different people that were used before to perform this format on stage \cite{mathewson-ted-talks}.
These six presentations were presented in mixed order, so that the participants did not realize that there were two different sets of presentation sources.
Participants had to indicate whether they were presenting or listening to the presentation, and had to give two ratings, one for the given presentation as a whole, and one for the slide deck quality.
We used a five-level scale, with five being the highest score (\textit{Awful, Bad, Average, Good, Amazing}).
A disadvantage of this evaluation is that the performative nature might be a confounding variable.
However, this might also help the participants assess the quality of the slide deck better than rating the slide deck in separation.

\begin{figure}[ht!]
    \subfloat[]{%
      \includegraphics[width=0.35\textwidth]{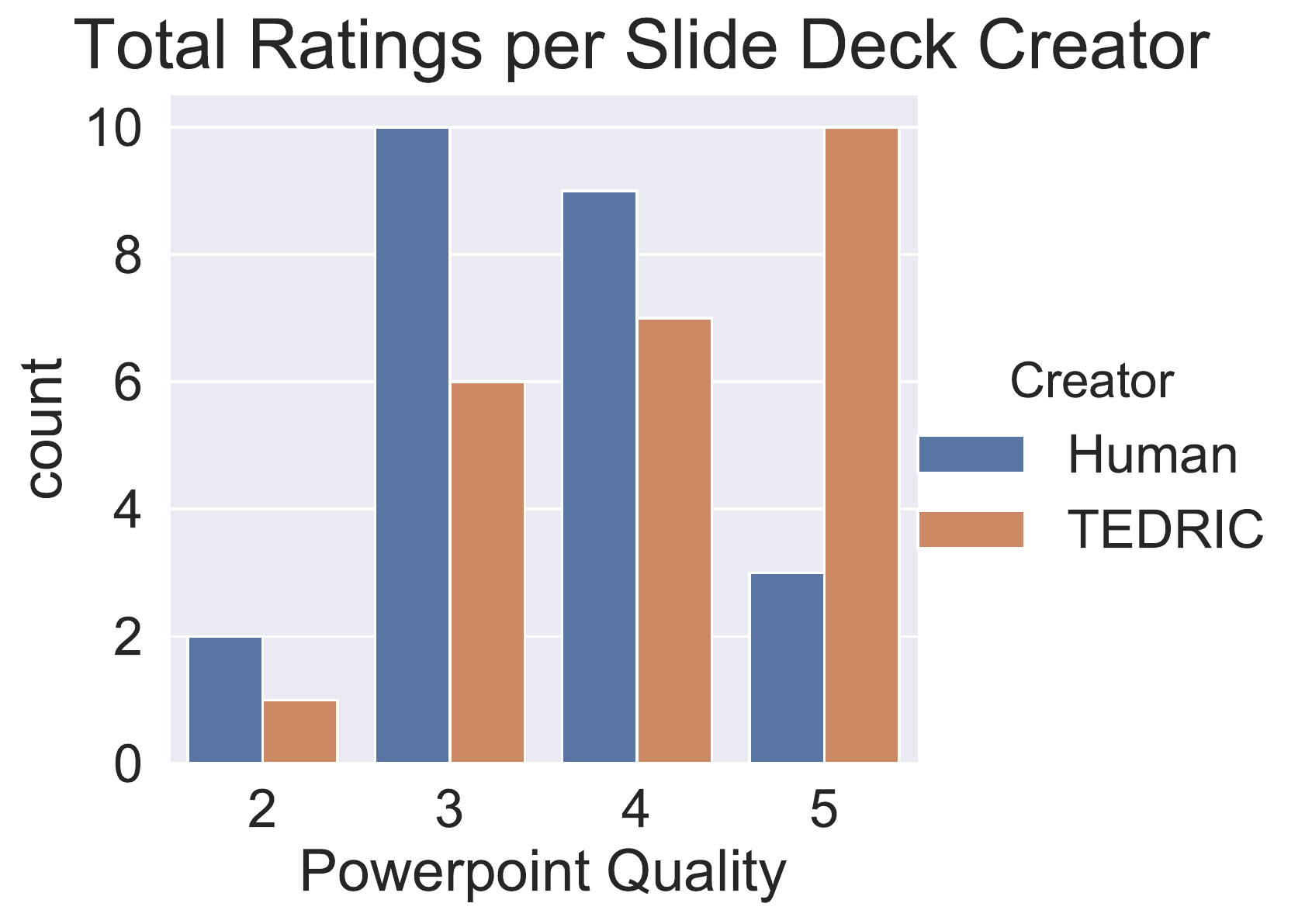}
     }
     \hfill
    \subfloat[]{
        \centering
        \begin{tabular}[b]{llll|l||lll|l}
            \toprule
                                    & \textbf{H1} & \textbf{H2} & \textbf{H3} & \textit{\textbf{Avg(H)}} & \textbf{G1} & \textbf{G2} & \textbf{G3} & \textit{\textbf{Avg(G)}} \\ \hline
        \textbf{Talk}    & 3.5         & 3.8        & 4.5         & \textit{3.9}            & 4.0         & 4.8         & 3.1         & \textit{4.0}            \\ \hline
        \textbf{Slide Deck} & 3.1         & 3.4         & 4.1         & \textit{3.5}            & 4.5         & 4.4         & 3.4         & \textit{4.1}    \\
            \bottomrule       
            
        \end{tabular}
    }

    \centering
    \caption{Comparing the (H)uman created and (G)enerated slide decks, for both the quality of the talks themselves and the slide deck quality (N=8, for all six presentations)}
    \label{fig:evaluation}
\end{figure}

With the gathered data\footnote{The evaluation data is made available on \url{https://github.com/korymath/tedric-analysis}} on Figure \ref{fig:evaluation}, we can check if the source of the slide deck is correlated with the quality of the slide deck.
When performing a t-test on the dataset to see if the distributions are different, we find $p=0.039$. This value allows us to reject, within the 5\% confidence interval, the hypothesis that they have the same distribution.
However, when performing a $\chi^2$ test on contingency table without scores of 1 (as none were present), then
$p=0.148$,
which is not significant enough to reject that they have a similar distribution.
This is likely due to the fact that there were only 48 observations.
Either way, the generated slides generally do not seem score lower in our sample than the human created slide decks.

\subsection{Qualitative Results}

\begin{figure}[ht!]
    \centering
    \includegraphics[width=\textwidth]{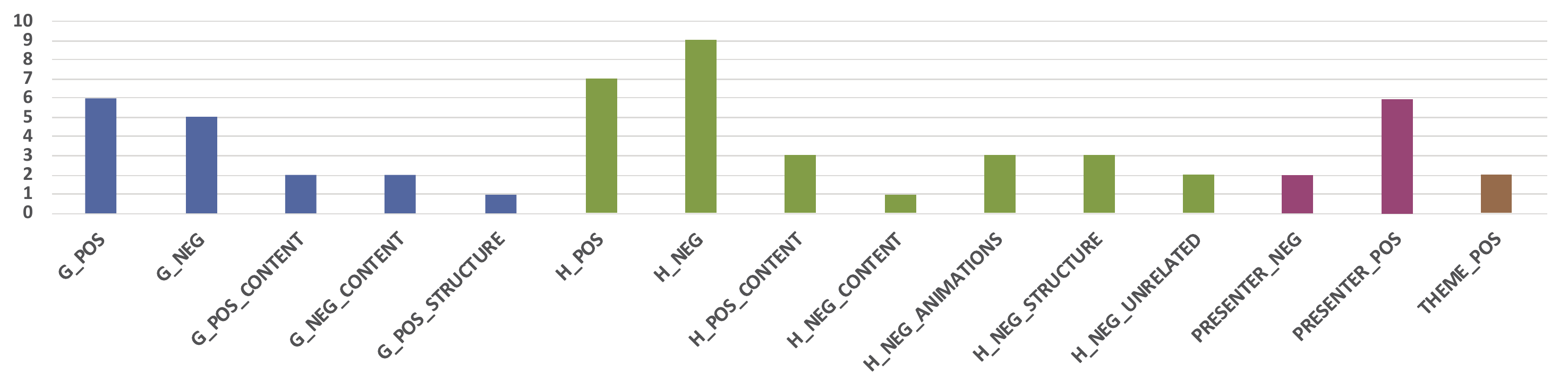}
    \caption{Codes and their frequency $f$ (if $f >0$) in the thematic analysis}
    \label{fig:thematic-analysis}
\end{figure}

The feedback forms also had a section where the participants could fill in short, textual comments about their given slide decks, of which they did not know the source.
There were 30 comments in total, which we described in a thematic analysis using codes that describe whether they are positive (POS) or negative (NEG) about the human created slide deck (H), generated slide deck (G), theme and presenter as well as common reasons for these sentiments (Figure \ref{fig:thematic-analysis}).

Some presenters had complained that it was quite unpredictable whether their given human-created slide decks would feature a new slide or an additional animation next, making it harder to present.
Some participants also complained that these human-created slide deck had nothing to do with the talk itself, and that it often put too many words in the presenter's mouth.
However, they did enjoy that slide decks had clever jokes, nice variation and backreferences with returning images.

Some participants complimented the generated slides on their nice structure and on the ease to link them to the presentation topic.
The participants wanted to try a variation after seeing five talks, and proposed to test out the format in a panel show setting, with one moderator and two guests.
The generated slide deck (\textit{G3} in Figure \ref{fig:evaluation}) did not support this type of talk well, as it was too focused on one person (e.g. in the \textit{``About Me''} slide), giving rise to some negative comments about the content.
The amount of text was too high, as there were three improvisers on stage, who all wanted to add new information as well.
In order to support this type of presentation exercise well in the future, we will thus need an additional, revised version of the presentation schema.

\subsection{Time Performance}

Since our generator is used in front of a live audience, it should be able generate within a reasonable time frame.
To optimally estimate the timings of the typical use case of this generator, we mimicked such a situation as well as possible.
Firstly, we used 500 of the most common English nouns\footnote{\url{https://www.wordexample.com/list/most-common-nouns-english}} as presentation topics for the generated slide decks, as these fairly accurately represent audience suggestions.
Choosing more common nouns as topics are also a worst-case estimate for time performance.
This is due to less common nouns having less relations, which increase the odds of slide topic generator cache hits, which also leads to more similar slide topics, increasing the probabilities of hitting content generator caches.
Secondly, we ran the timing experiment on a normal laptop with download speeds fluctuating between 80-150 megabits per second, as this is similar to the set-up a goal user might have.
Thirdly, the slide decks were not only represented in-memory, but also saved to a PowerPoint file.
Fourthly, we chose to generate slide decks of seven slides, as this is a nice length used for practising the format, as well as for presenting snappy, humorous presentations.
Increasing this length should not increase the generation time with a similar factor due to the fact that the slides are generated in parallel rounds.

\begin{figure}[ht!]
    \centering
    \includegraphics[width=\textwidth]{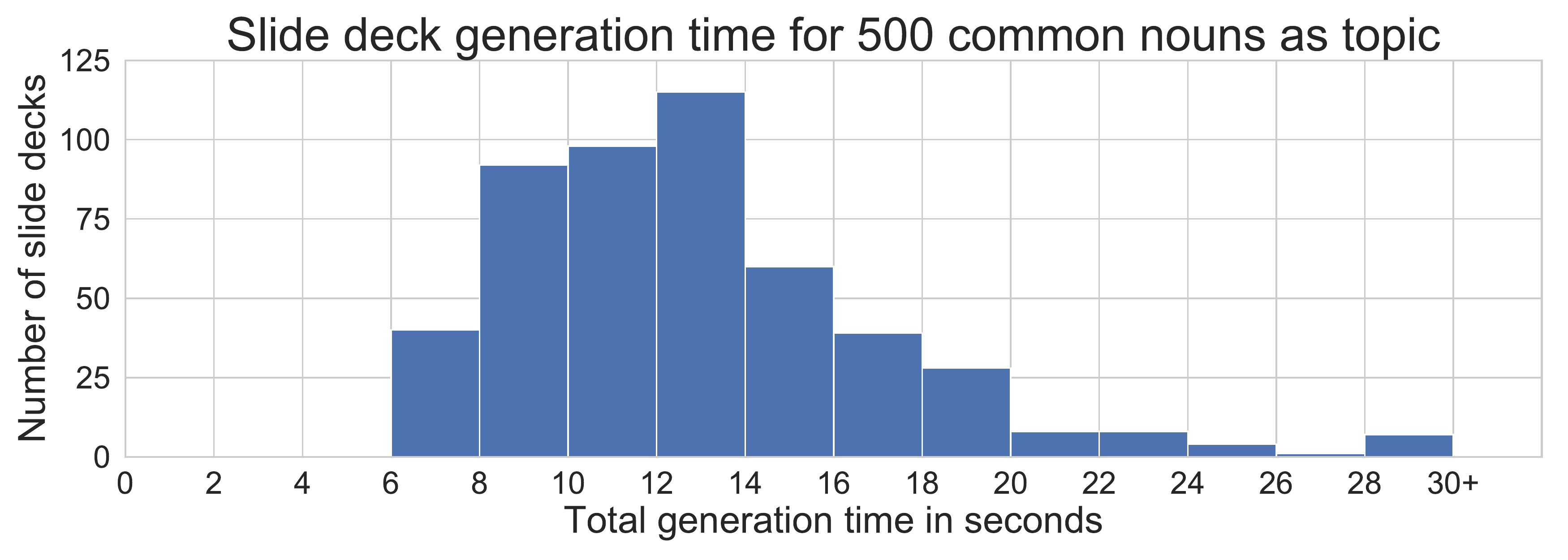}
    \caption{Generation time of slide decks with seven slides}
    \label{fig:timing_evaluation}
\end{figure}

We can see on Figure \ref{fig:timing_evaluation} that 94.4\% of slide decks are generated within a 20-second time window, with a median of 12.29 seconds, and a maximum of 38 seconds (and one outlier of 58 seconds).
The time spent on actual computations in the program has a median of 2.49 seconds, meaning that about 80\% of the total generation time is spent waiting on the response of remote content sources.
Better internet connections can thus decrease the generation time even further.
We can conclude from this experiment that the generation time of \system\ is low enough to be used in presentation workshops and on stage.

\subsection{Experiences on Exercise and On Stage Performance}

We organized a small workshop and coached a group of four improvisational theater actors using this generator.
They quickly improved their presenting methods.
The narrative of their stories initially lacked logic and did not link back to the generated title of the talk, or convinced the audience about the importance of the topic.
They also rarely dared using a bold statement right before going to the next slide, a technique for making the audience laugh (e.g. \textit{``The next slide will provide the solution to this problem.''}).
These skills were easily trained using the generator, and at the end of a three hour workshop, all participants gave talks that avoided these mistakes.
Later that week, they were put on stage during a comedy festival, where people could leave at any time to see other comedy acts.
They managed to keep all thirty to forty people interested enough to stay seated during all four talks.
Furthermore, they were able to make the audience laugh during almost every slide, thanks to the inspiring slide decks.
We see this as a type of collaborative creativity between humans and computers.
However, one thing we noticed was that the energy of the audience dropped when the fourth talk began.
A likely reason for this has been described in the past as an \textit{``ice cream after an ice cream situation''} \cite{mathewson-ted-talks}.
One way of overcoming this problem in the future would be to implement more types of presentation schemas, which could be achieved by slightly varying the current presentation schema (e.g. presentation schemas preferring more art, science or nature slides).

\subsection{Interpretation}
In the evaluation sample, the generated slide decks score higher on average than the human-created ones, and make the presenters give better presentations overall (Figure \ref{fig:evaluation}).
One explanation for this is that the generated presentations are more related to the given audience suggestions, whereas the premade slide decks are not.
This results in a more coherent presentation.
Even though this is a small evaluation sample, it is a good indication that the system is performing well and is usable for enhancing the performance of presenters.
We also showed that the generator is valuable and usable for coaching improvisers and even works well on stage.


\section{Future Work}


There are several extensions we propose for future work.
Obvious extensions are adding more variation to \system\ by adding more types of slide generators, different flavoured presentation schemas and slide deck themes.
We hope that making this tool available for other performers might also incentivize more tech-savvy users to add their new slide generators to the code repository, giving rise to such additional variations.

This research also opens the path to using genetic algorithms for more human-curated co-creation of slides.
This could be achieved by iteratively proposing several slides to the user, and giving the selected slide(s) higher fitness values.
A slide cross-over operator could use the slide generator to swap the content filled in certain placeholders with content from another slide.
Similarly, a mutation operator could just regenerate content in a certain position by calling this part of the generator again.
A user would then be able to make slides evolve using human curation, while the system keeps its important role in the creative aspects.
In this paper, we mainly put our focus on autonomous seeded slide generation, as the goal was to achieve real-time generation during performances.

One more ambitious goal of this research is to build an end-to-end system which can, in real-time, iteratively generate a slide deck
using interaction-based machine learning models \cite{knight-standup-robot,winters2018jokegeneration,jaques2018learning}.
This means that every next slide could be adapted to the response of the audience to previous slides, in order to further optimize engagement.
For these types of sequence generative models, the training data set must contain many examples of rich, contextual, structured 
pairs.
In order to generate such interactive presentations with e.g. neural models, we must develop an understanding of how to automatically generate a presentation comparable to a human using rules and heuristics, as we started in this paper.
We hope that our tool is used by performers who see the advantages of our system, as such a community could help us build a data set to open the path to a more interactive slide deck generation system.

\section{Conclusion}

In this paper, we presented a system for generating engaging slide decks called \system.
This system can be used for both training presentation skills as well as for performing improvisational comedy in the \textit{Improvised TED Talk} format.

There are several advantages our system has over humans having to create slide decks.
Firstly, it has the advantage of being able to create slide decks about an overarching theme decided by an audience suggestion.
We showed in our evaluation that this might increase the perceived quality of the given presentation.
Secondly, it performs the task of creating slide decks for this format several orders of magnitude faster than humans can.
Thirdly, it significantly lowers the barrier to perform improvised presentations.
Not only is it able to follow good guidelines, something that performers often have struggled to learn quickly before, it also relieves the performers of the time consuming task to create slide decks for each other to practise.

The behaviour of the presented system is a type of co-creation between computers and humans, as it provides surprising slides to spark the creativity of the presenter, similar to how humans creating a slide deck would.
We thus hope that this system will not only improve the presentation skills of humans, but also make them more acceptant towards computers aiding their creativity.

\subsubsection*{Acknowledgments}
Thank you to the reviewers for their time and attention reviewing this work, as well as for the insightful comments.
Thanks to Julian Faid and Dr. Piotr Mirowski for advice and support in the creation of the software.
Thank you to Lana Cuthbertson, the producer of TEDxRFT, and to the talented individuals at Rapid Fire Theatre for supporting innovative art.
Thank you to all the performers who have done improvised TED talks and shared their views and opinions on how to structure and frame, your help in building the design guide was critical.
Thanks to Shaun Farrugia for creating and hosting an online demo of \system , making it more easily available for performers of any background to use the system.
Thank you to volunteers from the Belgian improvisational comedy group Preparee for volunteering for the evaluation.

%
%
\bibliographystyle{splncs03}
\bibliography{winters}

\begin{thebibliography}{10}
\providecommand{\url}[1]{\texttt{#1}}
\providecommand{\urlprefix}{URL }

\bibitem{wilbur1981}
Wilbur, P.K.: Stand up, Speak up, or Shut up. New York: Dembner Books (1981)

\bibitem{doi:10.1080/08824096.2012.667772}
Dwyer, K.K., Davidson, M.M.: Is public speaking really more feared than death?
  Communication Research Reports  29(2),  99--107 (2012)

\bibitem{lee2014study}
Lee, S.: Study on the classification of speech anxiety using q-methodology
  analysis. Advances in Journalism and Communication  2(03), ~69 (2014)

\bibitem{watson2011perspective}
Watson, K.: Perspective: Serious play: teaching medical skills with
  improvisational theater techniques. Academic Medicine  86(10),  1260--1265
  (2011)

\bibitem{king1956comparison}
King, B.T., Janis, I.L.: Comparison of the effectiveness of improvised versus
  non-improvised role-playing in producing opinion changes. Human Relations
  9(2),  177--186 (1956)

\bibitem{mathewson-ted-talks}
Mathewson, K.W.: Improvised ted talks.
  \url{https://korymathewson.com/improvised-ted-talks/} (2018)

\bibitem{samim-ted-rnn}
Winiger, S.: Ted-rnn — machine generated ted-talks.
  \url{https://medium.com/@samim/ted-rnn-machine-generated-ted-talks-3dd682b894c0}
  (2015)

\bibitem{knight-standup-robot}
Knight, H.: Silicon-based comedy.
  \url{https://www.ted.com/talks/heather_knight_silicon_based_comedy} (2010)

\bibitem{mathewson2017improvised}
Mathewson, K.W., Mirowski, P.: Improvised theatre alongside artificial
  intelligences. In: AAAI Conference on Artificial Intelligence and Interactive
  Digital Entertainment (2017)

\bibitem{ppt-gen-academic}
Hu, Y., Wan, X.: Ppsgen: Learning-based presentation slides generation for
  academic papers. IEEE Transactions on Knowledge and Data Engineering  27(4),
  1085--1097 (April 2015)

\bibitem{ppt-gen-review}
Ranjan, A., Gangadhare, A., Shinde, S.V.: A review on learning based automatic
  ppt generation using machine learning. International Journal of Scientific
  Research in Computer Science, Engineering and Information Technology
  (IJSRCSEIT)  3,  547--552 (2018)

\bibitem{Sravanthi2009SlidesGenAG}
Sravanthi, M., Chowdary, C.R., Kumar, P.S.: Slidesgen: Automatic generation of
  presentation slides for a technical paper using summarization. In: FLAIRS
  Conference (2009)

\bibitem{PechaKucha}
Beyer, A.M.: Improving student presentations: Pecha kucha and just plain
  powerpoint. Teaching of Psychology  38(2),  122--126 (2011)

\bibitem{JAPE}
Binsted, K., Ritchie, G.: An implemented model of punning riddles. CoRR
  abs/cmp-lg/9406022 (1994)

\bibitem{STANDUP-Construction}
Manurung, R., Ritchie, G., Pain, H., Waller, A., Mara, D., Black, R.: The
  construction of a pun generator for language skills development. Applied
  Artificial Intelligence  22(9),  841--869 (2008)

\bibitem{venour:1999}
Venour, C.: {The computational generation of a class of puns}. Master's thesis,
  Queen's University, Kingston, Ontario (1999)

\bibitem{winters2018jokegeneration}
Winters, T., Nys, V., De~Schreye, D.: Automatic joke generation: Learning humor
  from examples. In: Streitz, N., Konomi, S. (eds.) Distributed, Ambient and
  Pervasive Interactions: Technologies and Contexts. pp. 360--377. Springer
  International Publishing, Cham (2018)

\bibitem{conceptnet}
Liu, H., Singh, P.: Conceptnet: A practical commonsense reasoning tool-kit. BT
  Technology Journal  22(4),  211--226 (Oct 2004)

\bibitem{tracery}
Compton, K., Kybartas, B., Mateas, M.: Tracery: An author-focused generative
  text tool. In: Schoenau-Fog, H., Bruni, L.E., Louchart, S., Baceviciute, S.
  (eds.) Interactive Storytelling. pp. 154--161. Springer International
  Publishing, Cham (2015)

\bibitem{jaques2018learning}
Jaques, N., Engel, J., Ha, D., Bertsch, F., Picard, R., Eck, D.: Learning via
  social awareness: improving sketch representations with facial feedback.
  arXiv preprint arXiv:1802.04877  (2018)

\end{thebibliography}
%

\end{document}